\documentclass[fleqn,usenatbib]{mnras} 
\usepackage{booktabs}
\usepackage{newtxtext,amsmath}
\usepackage{newtxmath}

\usepackage[T1]{fontenc} 

\DeclareRobustCommand{\VAN}[3]{#2}
\let\VANthebibliography\thebibliography
\def\thebibliography{\DeclareRobustCommand{\VAN}[3]{##3}\VANthebibliography}

\usepackage{graphicx}	
\usepackage{amsmath}	

\usepackage[mathlines]{lineno} 

\title[Disc-corona Evolution in AT 2019avd]{Evolution of accretion disc-corona in the TDE Candidate AT 2019avd}

\author[H. Xu et al.]{
Haichao Xu$^{1}$,
Xinwu Cao$^{1,2}$\thanks{E-mail: xwcao@zju.edu.cn (XC)},
Yanan Wang$^{3}$\thanks{E-mail: wangyn@bao.ac.cn (YW)},
and 
Andrzej A. Zdziarski$^{4}$
\\
$^{1}$Institute for Astronomy, School of Physics, Zhejiang University, 866 Yuhangtang Road, Hangzhou 310058, China\\
$^{2}$Center for Cosmology and Computational Astrophysics, School of Physics, Zhejiang University, 866 Yuhangtang Road, Hangzhou 310058, China\\
$^{3}$National Astronomical Observatories, Chinese Academy of Sciences, 20A Datun Road, Beijing 100101, China\\
$^{4}$Nicolaus Copernicus Astronomical Center, Polish Academy of Sciences, Bartycka 18, PL-00-716 Warszawa, Poland 
}

\date{Accepted 2025 October 28. Received 2025 October 21; in original form 2025 June 20}

\pubyear{\the\year{}} 

\begin{document}
\label{firstpage}
\pagerange{\pageref{firstpage}--\pageref{lastpage}}
\maketitle


\begin{abstract}  

X-ray observations of the tidal disruption event (TDE) candidate AT 2019avd show drastic variabilities in flux and spectral shape over hundreds of days, providing clues on the accretion disc-corona evolution.  
We utilize a disc-corona model,
in which a fraction of the gravitational energy released in the disc is transported into the hot corona above/below. Some soft photons emitted from the disc are upscattered to X-ray photons by the hot electrons in the optically thin corona.   
By fitting the \textit{NICER} observations of AT 2019avd during epochs when the spectra exhibit significant hardening, we derive the evolution of the mass accretion rate, $\dot{m}$, and the coronal energy fraction, $f$. 
Our results show that $f$ decreases with increasing $\dot{m}$, which is qualitatively consistent with that observed in active galactic nuclei (AGNs), while the slope of this source, $f\propto \dot{m}^{-0.30}$, is much shallower than that of AGNs. 
We also find that the non-thermal X-ray spectrum in this source is significantly softer than those typically seen in AGNs {and black-hole X-ray binaries}.
We argue that these quantitative differences can be a powerful diagnostic of the underlying magnetic turbulence, which may imply a stronger magnetic field within
the TDE accretion disc than that in typical AGNs.
It is also found that the evolution of the fitted neutral hydrogen column density follows a similar pattern to that of the accretion rate evolution, which may reflect the accumulation of absorbing material originating from the inflowing streams of stellar debris and/or other related sources.

\end{abstract}

\begin{keywords}
accretion, accretion discs – galaxies: active – X-rays: galaxies.
\end{keywords}



\section{Introduction}

Supermassive black holes (SMBHs) with masses ranging from $10^5$ to $10^{10}$ $M_\odot$ are believed to reside in almost all galaxies \citep[e.g.][]{kormendyCoevolutionNotSupermassive2013,heckmanCoevolutionGalaxiesSupermassive2014}. Through the accretion of surrounding gas, these black holes grow and convert gravitational energy into intense radiation, manifesting as active galactic nuclei (AGNs). In the standard AGN paradigm, the accreting material forms a geometrically thin, optically thick accretion disc in most bright AGNs \citep[e.g.][]{shakura1973black,frank2002accretion}. 
The characteristic temperature of the inner accretion disc is $\sim 10^4-10^5$~K, and the radiation from an accretion disc surrounding a SMBH is dominant in optical/UV wavebands \citep[e.g.][]{shakura1973black}, while the power-law hard X-ray emission has been detected ubiquitously in AGNs. 
It was suggested that the power-law hard X-ray spectra of AGN are most likely due to the inverse Compton scattering of soft photons emitted from the disc on a population of hot electrons in the corona above/below the disc \cite[e.g.][]{1979ApJ...229..318G,haardt1991two,svenssonBlackHoleAccretion1994}. In this disc-corona scenario, most gravitational energy is generated in the cold disc through the turbulence produced by the magnetorotational instability \citep[][]{1991ApJ...376..214B}. The magnetic fields generated in the cold disc may be strongly buoyant, and therefore a substantial fraction of magnetic energy is transported vertically to heat the corona above the disc with the re-connection of the fields \citep[][]{1998MNRAS.299L..15D,1999MNRAS.304..809D,2001MNRAS.328..958M,2002MNRAS.332..165M,caoAccretionDisccoronaModel2009}. 
The disc-corona model was extensively explored in many previous works, which is able to reproduce the main features of AGN spectra \citep[e.g.,][]{haardt1991two,1993ApJ...413..507H,svenssonBlackHoleAccretion1994,liuSimpleModelMagnetic2002,caoAccretionDisccoronaModel2009,2012ApJ...761..109Y}.
However, AGNs are fed by circumnuclear gas in quasi-steady state, and the timescale of AGN evolution is usually very long, which is comparable with the accretion timescale \citep[$\sim 10^{4-5}$~years; e.g.,][]{2021ApJ...916...61F}. It is therefore very difficult to monitor the spectral evolution of an individual AGN.

If the SMBH's tidal force dominates over the self-gravity of the star when it passes within the tidal radius of the SMBH, the star will be torn apart, resulting in a tidal disruption event (TDE; \citealt{hillsPossiblePowerSource1975,reesTidalDisruptionStars1988,phinneyManifestationsMassiveBlack1989}).The debris may form a disc surrounding the SMBH, and the light curve of the disc emission reaches a peak rapidly, and then decays on timescales of months to years. The released power of the disc can approach or even exceed the Eddington luminosity, which provides a unique opportunity to study the rapid evolution of accretion processes \citep[e.g.][]{evansTidalDisruptionStar1989a,gezariTidalDisruptionEvents2021}. 
Unlike AGNs, TDEs are powered by the sudden fallback of stellar debris, 
forming transient accretion systems. 
Many previous studies show strong evidence of compact accretion discs in TDEs \citep[e.g.][]{ lodatoMultibandLightCurves2011, piranDISKFORMATIONDISK2015,daiUnifiedModelTidal2018}. 
In several respects, these systems share similarities with black hole X-ray binaries (XRBs), such as rapid variability and the physics of the disc \citep[e.g.][]{remillardXRayPropertiesBlackHole2006}, while the mass feeding in XRBs is quite different from AGNs or TDEs \citep[][]{2019MNRAS.485.1916C}.

{X-ray emission of TDE can be described by a combination of a blackbody and a power-law component\citep[e.g.][]{KOMOSSA2015148}. The blackbody component with temperature in the range of $\sim 10^5-10^6$~K may probably be emitted from the disc, while the power-law component may originate in the corona.}
TDEs often exhibit extreme soft X-ray spectra in their early stages\citep[e.g.][]{badeDetectionExtremelySoft1996,komossaHugeDropXRay2004,KOMOSSA2015148}.
At the later stages, the thermal component tends to be weaken, and the X-ray emission becomes increasingly dominated by the power-law component with spectral hardening \citep[e.g.][]{badeDetectionExtremelySoft1996,komossaHugeDropXRay2004,KOMOSSA2015148,yaoMassiveBlackHole2025}.
These findings provide useful information of the evolution of transient disc-corona systems in TDEs. 
{In recent years, a few studies have begun exploring the application of disk-corona models to TDEs \citep[e.g.][]{10.1093/mnras/stac3278,mageshwaranDiskcoronaModelingSpectral2023}.}

The high-energy nuclear transient AT 2019avd, located at a redshift of $z=0.028$, was first detected in the optical band by the Zwicky Transient Facility \citep[ZTF,][]{bellmZwickyTransientFacility2018} on 2019, February 9 \citep[][]{nordinZTFTransientDiscovery2019}. Its significant flaring activities have been captured by multi-wavelength follow-up observations \citep[e.g.][]{malyali2019avdNovelAddition2021,wangRadioDetectionAccretion2023}. AT 2019avd exhibits several characteristics consistent with a TDE, including ultra-soft X-ray spectra and optical spectral lines typical of known TDEs. 
Although its double-peaked optical light curve is somewhat unusual, similar optical evolution has been observed in other TDEs \citep[e.g.][]{yaoTidalDisruptionEvent2023}. Thus, it is classified as a TDE candidate. Although the precise trigger of the X-ray activity relative to the optical outburst remains uncertain, several studies suggest that the X-ray flare occurred after the initial optical brightening \citep[e.g.][]{malyali2019avdNovelAddition2021, chen2019avdTidalDisruption2022, wangRadioDetectionAccretion2023}.

\citet{wangRapidDimmingFollowed2024} reported intensive X-ray monitoring of AT 2019avd with the \textit{NICER}, \textit{Swift} and \textit{Chandra} telescopes between 2020 September 19 and 2021 June 16. Based on the X-ray properties, they divided this period into five phases. Setting MJD 59110 as Day 0—the moment when AT 2019avd reached its peak X-ray luminosity—the phases correspond to Day 0-100 (Phase 1), Day 101-172 (Phase 2), Day 173-225 (Phase 3), Day 226-249 (Phase 4), and Day 250 onward (Phase 5). In Phase 1, the X-ray luminosity declined by about a factor of five with little variations in the hardness ratio (defined as the photon count ratio between the 0.8--2 keV and 0.3--0.8 keV bands). In Phase 2, the luminosity unexpectedly rebrightened to a secondary peak. Phase 3 saw a rapid luminosity drop by approximately an order of magnitude, followed by further fading during Phases 4 and 5, where AT 2019avd became extremely faint in X-rays. Throughout Phases 2-5, the X-ray hardness ratio steadily increased. This evolution pattern suggests the possible evolution of a disc-corona system.

In this work, we adopt an accretion disc-corona model developed by \cite{caoAccretionDisccoronaModel2009}, which computes the X-ray continuum spectrum of the corona for specified parameters, to describe the late-time X-ray hardening of AT 2019avd.
By fitting the \textit{NICER} X-ray spectra of AT~2019avd during Phases 3 and 4 with our model,  
we derive the temporal evolution of the accretion rate, the fraction of gravitational energy dissipated in the corona, and the coronal temperature.
The framework of the disc-corona model, and the results are described in 
in Sections \ref{sec:model} and \ref{sec:results} respectively,  
In Section~\ref{sec:fitting}, we fit the observational data of AT2019 avd during Phases 3 and 4 using the model. Section \ref{sec:discussion} 
contains the discussion.

\section{The disc-corona model}
\label{sec:model}

We consider a system composed of a geometrically thin and optically thick accretion disc and a slab-like, geometrically thick and optically thin hot corona situated above and below the disc. The gravitational energy of the gas is released in the disc through viscous processes, which is believed to be the turbulence most probably triggered by the magneto-rotational instability \citep[][]{1991ApJ...376..214B}.   
A fraction of released gravitational energy is transported vertically into the corona by magnetic fields of the disc, which heats the hot plasma in the corona via magnetic reconnection, while the remainder is radiated out from the disc. The heating of the corona is balanced by radiative cooling, which is dominated by the inverse Compton scattering of soft photons from the underlying disc by the hot coronal electrons.

In our model, we neglect the spin of the black hole and adopt the standard accretion disc framework \citep[e.g.][]{shakura1973black,frank2002accretion}. The gravitational energy released by viscous dissipation per unit area is given by
\begin{equation}
Q^+_\text{dissi} = \frac{3G\dot{M}M_\text{BH}}{8\pi R^3}\left[1 - \left(\frac{R_{\rm in}}{R}\right)^{1/2}\right],
\label{eq:accretion}
\end{equation}
where $\dot{M}$ is the mass accretion rate and $R_{\rm in}$ is the inner radius of the accretion disc.

Approximately half of the hard photons are reflected back onto the accretion disc, and most of them are absorbed by the disc \cite[e.g.][]{haardt1991two}. The reflection albedo $a$ for the hard photons by the disc is typically low 
\citep[e.g.][]{zdziarskiCorrelationComptonReflection1999}. We adopt a typical value of a=0.2 in our work, as suggested by  
\citet{haardt1991two,vasudevanPiecingTogetherXray2007}. 
The energy equation for the accretion disc is 
\begin{equation}
Q^+_\text{dissi} - Q^+_\text{cor} + \frac{1}{2}(1-a)Q^+_\text{cor}= \sigma_\text{sb} T^4_\text{s},
\label{eq:partition}
\end{equation}
where $\sigma_\text{sb}$ is the Stefan-Boltzmann constant, and $T_\text{s}$ is the effective blackbody temperature of the accretion disc.

The transfer rate of energy is defined as the ratio of the heating rate of the corona to the viscous dissipation rate in the cold disc, i.e.
\begin{equation}
f = {Q^+_\text{cor}}/{Q^+_\text{dissi}}.
\label{eq:transfer}
\end{equation} 
which, in principle, can be calculated based on a suitable magnetic field generation model in the disc by assuming the corona is heated by re-connection of the field \citep[][]{1998MNRAS.299L..15D,1999MNRAS.304..809D,2001MNRAS.328..958M,2002MNRAS.332..165M,caoAccretionDisccoronaModel2009}. 
However, the transfer of the energy from the disc to corona is highly dependent on the the physical processes of magnetic field generation, buoyancy, in the disc, and its eventual dissipation in the corona. In the work of  \cite{caoAccretionDisccoronaModel2009}, for instance, the buoyant magnetic field is derived using an angular momentum equation that includes a specific magnetic stress tensor, which is then used to calculate the power dissipated in the corona and thus the value of $f$. Since the details of these magnetic processes remain highly uncertain, and to avoid being constrained by a particular unverified model, we treat $f$ as a free input parameter in this work, which is similar to some previous works \citep[e.g.][]{haardt1991two, 1993ApJ...413..507H, svenssonBlackHoleAccretion1994}. Consequently the explicit angular momentum equation is not required for our calculations.

The equation of state for the plasma in the corona can be expressed as
\begin{equation}
p = \sum_\text{j} n_\text{j} k T_\text{j} + n_\text{e} k T_\text{e} + p_\text{mag},
\label{eq:state}
\end{equation}
where the summation is over all ion species and $p_\text{mag}$ is the magnetic pressure in the corona. We assume that the gas consists of $3/4$ hydrogen and $1/4$ helium by mass, such that $n_\text{H}=12n_\text{e}/14$ and $n_\text{He}=n_\text{e}/14$. Furthermore, the temperatures of hydrogen and helium ions are assumed to be equal, i.e., $T_\text{H} = T_\text{He} = T_\text{i}$. In this work, the magnetic pressure is assumed to remain equal to the gas pressure \citep[e.g.][]{liuSimpleModelMagnetic2002,liuSpectraMagneticReconnectionheated2003,caoAccretionDisccoronaModel2009}. The sound speed in the corona and the corresponding vertical scale height are defined as
\begin{equation}
c_\text{s} = \left(\frac{p}{\rho}\right)^{1/2} = \left[(1+\beta)\left(\frac{kT_\text{i}}{\mu_\text{i} m_\text{p}}+\frac{kT_\text{e}}{\mu_\text{e} m_\text{p}}\right)\right]^{1/2},
\label{eq:sound}
\end{equation}
and
\begin{equation}
H = \frac{c_\text{s}}{\Omega_\text{K}} = \left[\frac{R^3(1+\beta)}{GM_\text{BH}}\left(\frac{kT_\text{i}}{\mu_\text{i} m_\text{p}}+\frac{kT_\text{e}}{\mu_\text{e} m_\text{p}}\right)\right]^{1/2},
\label{eq:scale}
\end{equation}
respectively, where $\beta$ is the ratio of the magnetic pressure to the gas pressure and $\beta=1$ is adopted in the calculations and $\mu_\text{i}=1.23,\mu_\text{e}=1.14$ are the mean molecular weights of ions and electrons, respectively.

The two-temperature plasma is cooling predominantly through electron radiation due to the significant mass difference between ions and electrons. The energy transferred from ions to electrons via Coulomb collisions \citep[e.g.][]{stepneyNumericalFitsImportant1983a}. It was also suggested that the strong magnetic fields in the corona can efficiently heat the electrons through magnetic reconnection \citep[e.g.][]{1979ApJ...229..318G,bisnovatyi-koganInfluenceOhmicHeating1997,bisnovatyi-koganMagneticFieldLimitations2000}. Thus, the energy equation for the corona can be written as
\begin{equation}
Q^+_\text{cor} = Q_\text{ie} + \delta Q^+_\text{cor},
\label{eq:corona}
\end{equation}
where $\delta$ is a parameter that accounts for the additional heating ratio due to magnetic reconnection. In this work, we adopt $\delta = 0.5$ \citep[e.g.][]{caoAccretionDisccoronaModel2009}. 

We assume that ions and electrons in the corona are always in thermal equilibrium, following a relativistic Maxwellian distribution developed by \cite{juttner1911maxwellsche}:
\begin{equation}
N(\gamma) = \frac{\gamma^2 \beta}{\theta_\text{e} K_2(1/\theta_\text{e})} \exp\left(-\frac{\gamma}{\theta_\text{e}}\right),
\label{eq:relativistic }
\end{equation}
where $\theta_\text{e} = kT_\text{e}/{m_\text{e}c^2}$ is the dimensionless temperature of electrons. The Coulomb collision cooling rate per unit area is given by 
\begin{equation}
\begin{aligned}
&Q_{\text{ie}}=\sum_\text{j}\frac{3}{2}\frac{m_\text{e}Z_\text{j}^2}{m_\text{p} A_\text{j}}n_\text{e}n_\text{j}H\sigma_\text{T}c\frac{k(T_\text{i}-T_\text{e})}{K_2(1/\theta_\text{e})K_2(1/\theta_\text{j})}\log\Lambda \\
&\times\left[\frac{2(\theta_e+\theta_\text{j})^2+1}{\theta_e+\theta_\text{j}}K_1\left(\frac{\theta_e+\theta_\text{j}}{\theta_e\theta_\text{j}}\right)+2K_0\left(\frac{\theta_e+\theta_\text{j}}{\theta_e\theta_\text{j}}\right)\right],
\end{aligned}
\label{eq:ie}
\end{equation}
where $\theta_\text{j} = {kT_\text{i}}/{A_\text{j}m_\text{p}c^2}$ are the dimensionless temperatures of the $\text{j}-$th ion species, and $A_\text{j}$ and $Z_\text{j}$ are the mass and charge numbers of the ions, respectively. In this work, $\log\Lambda=20$ is adopted \citep[e.g.][]{stepneyNumericalFitsImportant1983a,zdziarskiHotAccretionDiscs1998,caoAccretionDisccoronaModel2009}.

The emission of synchrotron and bremsstrahlung radiation of the corona is always small, which is negligible compared to radiation due to inverse Compton scattering \citep[e.g.][]{caoAccretionDisccoronaModel2009,schnittmanXRAYSPECTRAMAGNETOHYDRODYNAMIC2013}. However, bremsstrahlung emission may contribute significantly in the hard X-ray band, which may set a constraint on the electron temperature $T_{\rm e}$ in the corona with observations. So we include both these two in the energy equation of the corona 
\citep[e.g.][]{kawaguchiBroadbandSpectralEnergy2001}, i.e.,
\begin{equation}
    Q^+_\text{cor} = Q^-_\text{comp} + Q^-_\text{br}.
    \label{eq:cooling}
\end{equation}

In the cases of our interest, $\theta_\text{e}$ is always less than 1, so electron-electron bremsstrahlung can be neglected compared to electron-ion bremsstrahlung.  According to \citet{betheStoppingFastParticles1934,stepneyNumericalFitsImportant1983a,heitler1984quantum}, the emissivity of thermal bremsstrahlung is given by
\begin{equation}
    \chi_\omega=\frac{dE}{dVdt\,d\omega} =\sum_{\text{j}} Z_j^2 n_j n_e c \int_{1+\omega}^{\infty} \omega \frac{d\sigma}{d\omega} \beta N(\gamma) d\gamma,
    \label{eq:bremsstrahlung_rate}
\end{equation}
where $\omega=h\nu/m_\text{e}c^2$ is the dimensionless photon energy, and $d\sigma/d\omega$ is the Bethe-Heitler cross-section. 
\cite{manmotoSpectrumOpticallyThin1997} solved the radiative transfer equation for parallel slab and found the bremsstrahlung spectrum emitted per unit surface area can be expressed as:
\begin{equation}
    F_\text{br}(\nu)=\frac{2\pi}{\sqrt{3}} B_{\nu} \left[1 - \exp\left(-2\sqrt{3}\tau_{\nu}^{*}\right)\right],
\end{equation}
where $B_{\nu}$ is the Planck function, and $\tau_{\nu}^{*}=\sqrt{\pi}H\chi_\nu/2$. The total bremsstrahlung cooling rate $Q_{br}^{-}$ in Equation (\ref{eq:cooling}) is then obtained by integrating $F_{br}(\nu)$ over all frequencies.

For inverse Compton scattering, we follow \cite{coppi1990reaction} and \cite{kino2000radiation} to compute the radiation spectrum. The spectrum for the $n$-th scattering event is given by:  
\begin{equation}
    F_{n}(\omega) = \int_{\gamma} \int_{\omega'} R(\omega', \gamma) P(\omega;\omega',\gamma) N(\gamma) F_{n-1}({\omega'}) d\gamma d\omega',
\end{equation}
where $R(\omega, \gamma)$ is the scattering rate between photons and electrons and $P(\omega;\omega',\gamma)$ is the energy distribution of the scattered photon. 
In our calculations, we approximate $P(\omega;\omega',\gamma)$ using a delta function $\delta\left[\omega-\langle\omega(\omega',\gamma)\rangle\right]$.
The total inverse Compton spectrum is given by:
\begin{equation}
    F_\text{comp}(\nu) = \sum_{n=1}^{\infty} P_n(\tau) F_{n}(\nu),
    \label{eq:compton_spectrum}
\end{equation}
where $\tau = \sigma_\text{T} n_\text{e} H$ is the electron scattering optical depth of the corona and $P_n(\tau) = e^{-\tau}{\tau^n/n!} $ is the Poisson distribution for the probability distribution of photon-electron scattering events \citep[e.g.][]{esinHotOneTemperatureAccretion1996,kino2000radiation}.   

In the context of our disc-corona system, the incident photon spectrum $F_{\rm in}(\nu)$ is given by the thermal radiation from the underlying accretion disc and the thermal bremsstrahlung radiation, which is generated within the corona itself, i.e.
\begin{equation}
    F_{\rm in}(\nu) = F_0(\nu) = F_{\rm disc}(\nu) + F_{\rm br}(\nu).
\end{equation}
And the total inverse Compton scattering cooling rate $Q^-_\text{comp}$ is given by
\begin{equation}
    Q^-_\text{comp} = \int_{0}^{\infty} F_\text{comp}(\nu) +\left(e^{-\tau}-1\right)F_\text{in}(\nu) d\nu.
\end{equation}

It has been found that the radiation emitted from the surface of the accretion disc is not a perfect blackbody spectrum. The presence of heated gas at the disc's surface results in radiation in the high-energy bands that exceeds the blackbody radiation \citep[e.g.][]{hubenyNonLTEModelsTheoretical2001}. \cite{chiangXRayReprocessingModel2002a} proposed a temperature-dependent frequency correction factor, i.e., a hardening factor,
\begin{equation}
    f_{\rm col}(T_{\rm s}) = f_{\infty} - \frac{(f_{\infty} - 1)[1 + \exp(-\nu_\text{b} / \Delta\nu)]}{1 + \exp\left[(\nu_\text{p} - \nu_\text{b}) / \Delta\nu\right]},
    \label{eq:col_correction}
\end{equation}
where $f_\infty=2.3,\nu_\text{p}=2.82kT_{\rm s}/h$, and $\nu_\text{b}=\Delta \nu=5\times10^{15}\text{Hz}$. The emission spectrum of the accretion disc is then given by \cite{chiangXRayReprocessingModel2002a,caoJetFormationBL2003a}
\begin{equation}
    F_{\text{disc}}(\nu) = \frac{2\pi h \nu^3/f_{\text{col}}^4}{c^2\left[\exp(h\nu/f_{\text{col}}k T_{\text{s}}) - 1\right]}.
    \label{eq:disk}
\end{equation}

We also need to know the temperature of the ions in the corona. Similar to \cite{caoAccretionDisccoronaModel2009}, we set the ion temperature to 0.9 times the virial temperature, i.e.,
\begin{equation}
    T_\text{i} = 0.9 T_\text{vir} = 0.9\frac{GM_\text{BH}m_\text{p}}{3kR}.
\end{equation}
We find that the final results of the model are insensitive to the values of $T_{\rm i}$.

For given values of accretion rate $\dot{M}$ and coronal energy fraction $f$, the structure of the corona are determined by solving the energy balance equations, i.e., Equations (\ref{eq:partition}), (\ref{eq:transfer}), (\ref{eq:corona}), and (\ref{eq:cooling}).
The radiation spectrum of the disc-corona system is given by
\begin{equation}
    F_\text{tot}(\nu)=e^{-\tau}F_\text{disc}(\nu)+\frac{1+a}{2}F_\text{comp}(\nu)+(1+a)F_\text{br}(\nu),
\end{equation} 
in which the three terms in the right side represent the radiation from the accretion disc without being  scattered in the corona that can reach the observer, the inverse Compton scattering component from the corona, and the bremsstrahlung component from the corona, respectively. In the first term, we only consider the scattering of disc photons, as absorption is negligible compared to scattering in the hot, tenuous plasma. In the second term, we assume that approximately half of the Comptonized photons are scattered upwards and are directly observed, while the other half are scattered downwards towards the disc. A fraction of the downward radiation is then reflected by the disc and reaches the observer. For simplicity, we assume that the spectral shape of the reflected component is similar to the incident one. Similar to the second term, the third term effectively includes photons that travel directly outwards as well as those reflected from the disc, under the assumption that reflection does not significantly alter the spectral shape.

\section{Results of the model}
\label{sec:results}

In this work, we set the inner radius of the accretion disc to the innermost stable circular orbit (ISCO) for a Schwarzschild black hole, i.e., $R_{\rm in} = 6 R_{\rm g}$, where $R_\text{g} = GM_\text{BH}/c^2$ is the gravitational radius of the black hole.  
The tidal radius for a solar-like star being disrupted by a SMBH with $\sim 10^6~M_\odot$ is approximately $R_{\rm T}\sim 50 R_{\rm g}$ \citep[e.g.][]{hillsPossiblePowerSource1975, hayasakiCircularizationTidallyDisrupted2016}. For such a star in a nearly parabolic orbit, the self-interaction of the resulting debris stream is expected to be  rapidly circularized, which leads to the formation of a disc with a characteristic size of $R_{\rm C}\approx 2 R_{\rm T}$ \citep[e.g.][]{shiokawaGENERALRELATIVISTICHYDRODYNAMIC2015, bonnerotDiscFormationTidal2016, hayasakiCircularizationTidallyDisrupted2016}. Therefore, for simplicity, we adopt a fiducial outer radius of $R_{\rm out}=100 R_{\rm g}$ in our calculations. We note that the results are quite insensitive to the outer radius of the disc, because most gravitational energy of the accreting gas is released in the inner region of the disc near the BH.

\begin{figure} 
    \centering
    \includegraphics[width=\columnwidth]{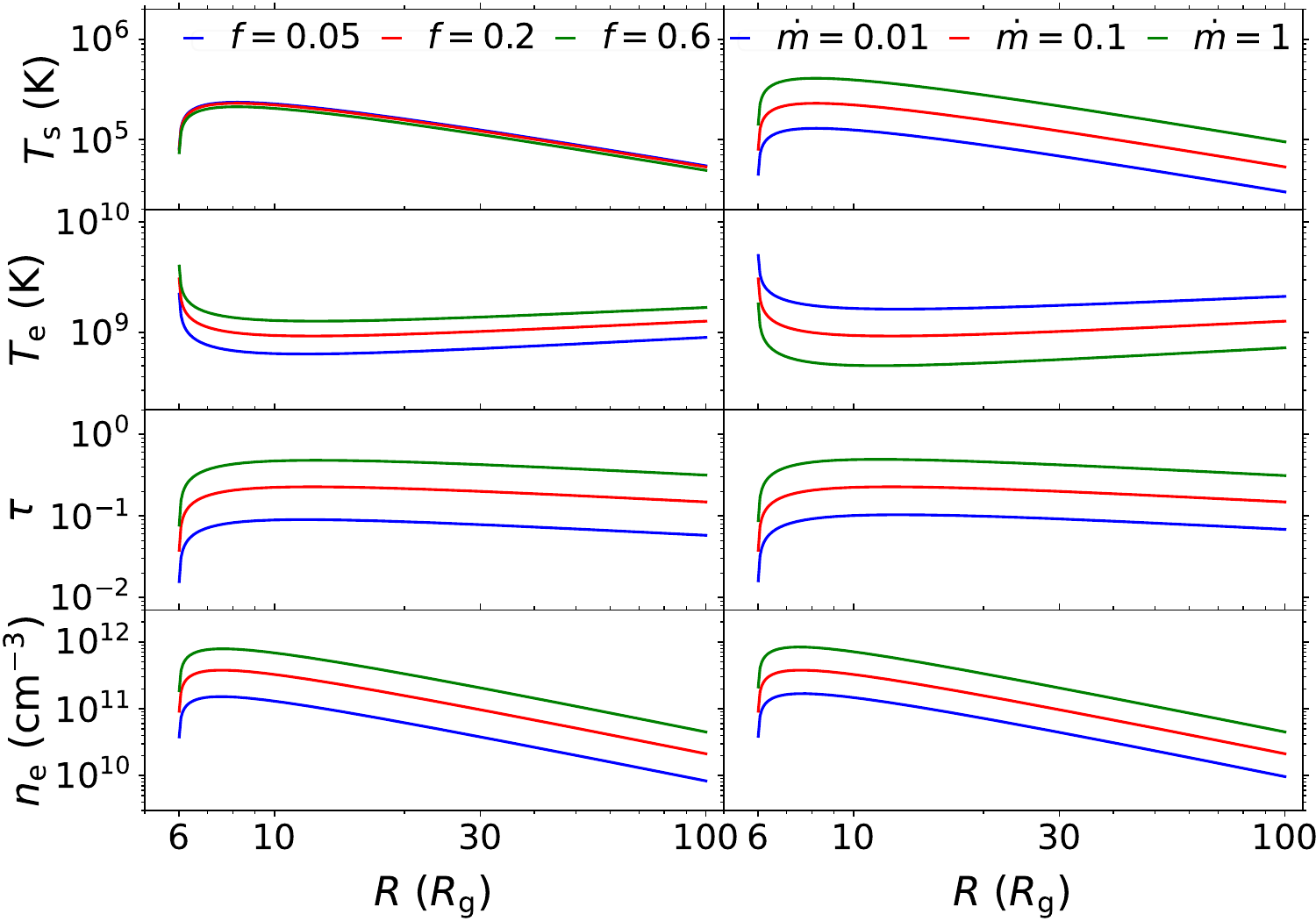}
    \caption{Radial profiles of the disc-corona system. The left column of panels shows  the effect of varying the coronal fraction $f$ at a fixed accretion rate of $\dot{m}=0.1$. The right column shows the effect of varying the accretion rate $\dot{m}$ at a fixed coronal fraction of $f=0.2$. From top to bottom, the panels show: the effective temperature of the accretion disc, the electron temperature, the optical depth, and the electron number density of the corona. A black hole mass of $M_{\rm BH} = 10^6 M_\odot$ is adopted.}
    \label{fig:structure}

\end{figure}

\begin{figure}
    \centering
    \includegraphics[width=\columnwidth]{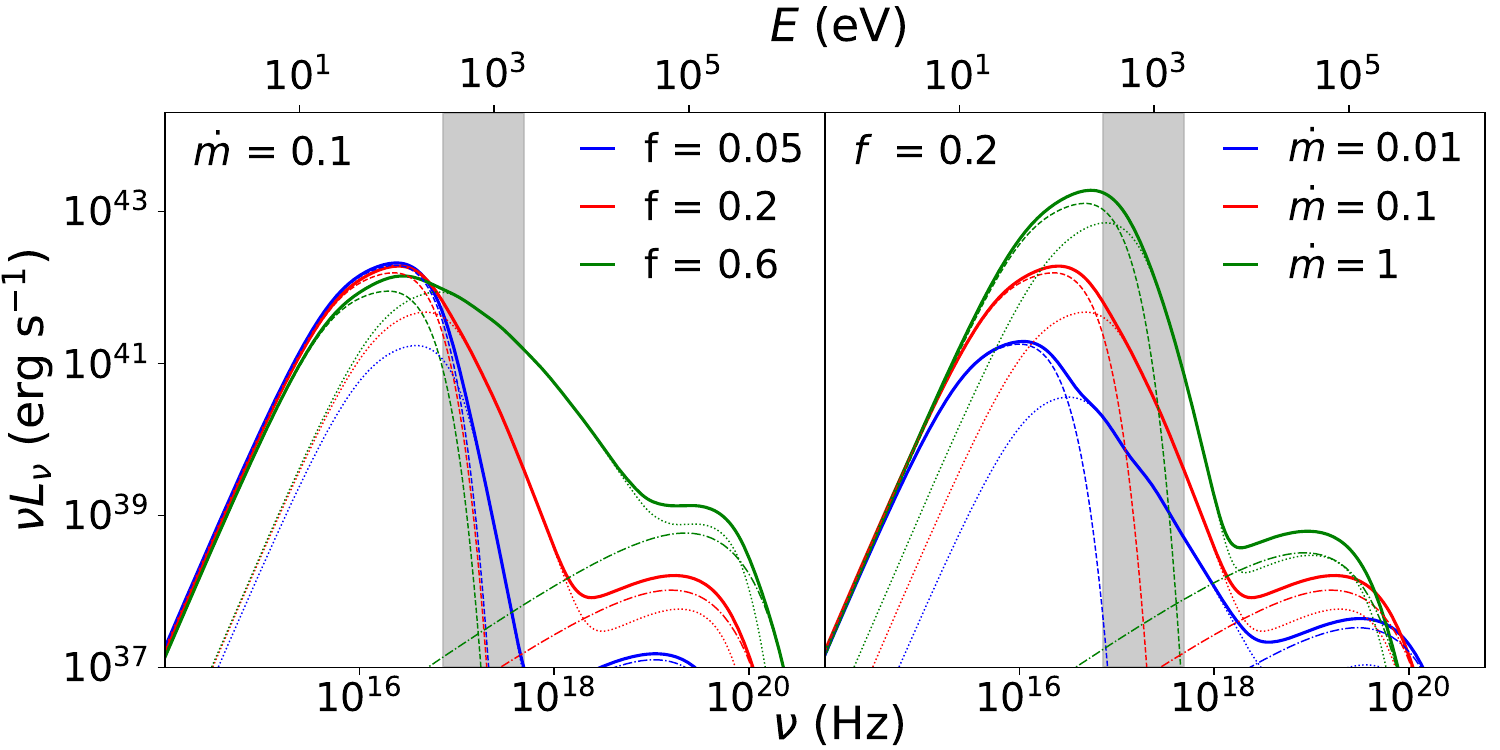}
    \caption{The spectra of the disc-corona systems for different accretion rates $\dot{m}$ and coronal fractions $f$. The dashed lines indicate the modified disc emission, while the dash-dotted and dotted lines represent inverse Compton scattering and thermal bremsstrahlung components of the corona, respectively. Grey shaded regions highlight the 0.3--2~keV X-ray bands. A black hole mass of $M_{\rm BH} = 10^6 M_\odot$ is adopted. }
    \label{fig:spectrum}
\end{figure}

It was found that the radial energy advection in the accretion disc may be important when the accretion rate is significantly higher than the Eddington rate \citep[][]{abramowicz1988slim}. For a slim accretion disc accreting at the Eddington rate without a corona, the fraction of the radially advected power to the viscously dissipated power is $\lesssim 0.2$ \citep[][]{2022ApJ...936..141C}. For the present work, a fraction of the dissipated power in the disc is tapped into the corona, which means the fraction of radially advected power should be lower than that given for a slim disc without a corona. Therefore, we focus on the case of sub-Eddington or slightly super Eddington accretion, $0.01<\dot{m}<2$, where $\dot{m}= 0.1\dot{M} c^2/L_{\rm Edd}$ ($L_{\rm Edd}$ is the Eddington luminosity). The structure of the corona can be calculated when the values of the parameters $\dot{m}$ and $f$, and the BH mass are specified. The structures of the coronae are given in Figure \ref{fig:structure} for typical values of the disc parameters. With the derived corona structures, we can calculate the spectra of the disc-corona systems accordingly, which are given in Figure \ref{fig:spectrum}.

Based on the model introduced in Section \ref{sec:model}, we computed the structure and emission spectra of the disc-corona system for a black hole mass of $M_{\rm BH}=10^{6}M_{\odot}$ and a range of accretion rates $\dot{m}$ and coronal energy fractions $f$. Figure \ref{fig:structure} shows the effective temperature profile of the accretion disc and the thermodynamic structure of the corona. We find that the effective temperature of the accretion disc $T_{\rm s}$ increases mainly with increasing $\dot{m}$, while its dependence on $f$ is negligible. Therefore, in the optical/UV bands, the disc emission in our disc-corona system is modified only by a factor of $e^{-\tau}$ compared with the situation where only the disc is present.

As shown in Figure~\ref{fig:structure}, the electron temperature $T_{\rm e}$ in the corona remains significantly below the local virial temperature, which is approximately $6\times 10^{11}{\rm K}$ in the inner disc region ($R\sim 6R_{\rm g}$) and decreases to $\sim4\times10^{10}$K near the outer disc radius ($R\sim 100 R_{\rm g}$). According to Equations (\ref{eq:sound}) and (\ref{eq:scale}), this leads to a scale height that increases nearly linearly with radius, i.e., $H\propto R$. the electron temperature $T_{\rm e}$ and optical depth $\tau$ exhibit only mild radial variation across the corona. The electron number density $n_{\rm e}$ decreases roughly as $R^{-1}$, consistent with the near-constant optical depth. Compared to the disc's effective temperature, the coronal parameters are more sensitive to variations in both $\dot{m}$ and $f$.

\section{Evolution of the accretion disc-corona system in AT 2019avd}
\label{sec:fitting}

In our analysis of AT 2019avd, we adopt some system parameters from previous studies, including a redshift of $z = 0.028$ \citep[e.g.][]{malyali2019avdNovelAddition2021}, a luminosity distance of $D = 130 \text{ Mpc}$ \citep[e.g.][]{wangRadioDetectionAccretion2023}, and a black-hole mass of $M_{\rm BH} = 10^{6.3} M_\odot$ \citep[e.g.][]{malyali2019avdNovelAddition2021}. The hydrogen column density is constrained to the range $N_{\rm H}=(2.4-7.0)\times10^{20}\text{cm}^{-2}$, based on Galactic measurements by HI4PI Collaboration\cite[e.g.][]{bekhtiHI4PIFullskySurvey2016} and spectral modelling from \cite{wangRapidDimmingFollowed2024}.

To ensure compatibility with our disc-corona framework, we focus on the post-peak evolution of AT 2019avd following its second X-ray maximum. Specifically, the dataset includes the majority of X-ray spectra from Phase 3 and all spectra from Phase 4, as defined by \cite{wangRapidDimmingFollowed2024}, covering the transition from near- to sub-Eddington accretion regimes. During these intervals, the optical/UV luminosity remains consistently low. We use these spectra to probe the structural and radiative evolution of the disc-corona system as the source transitions from high to moderate accretion rates.

As reported by \citet{wangRapidDimmingFollowed2024}, in addition to AT 2019avd, three other known X-ray sources, i.e. 1RXS J082334.6+042030, 2MASX J08232985+0423327, and IC 505, are present in the \textit{NICER} field of view. The contributions from 1RXS J082334.6+042030 and 2MASX J08232985+0423327 are negligible during the period studied in this work. For IC 505, \citet{wangRapidDimmingFollowed2024} created a spectral template to represent its contribution (for further details, see Appendix A of \citet{wangRapidDimmingFollowed2024}). We adopted this template in our fit.

\begin{figure}
    \centering
    \includegraphics[width=\columnwidth]{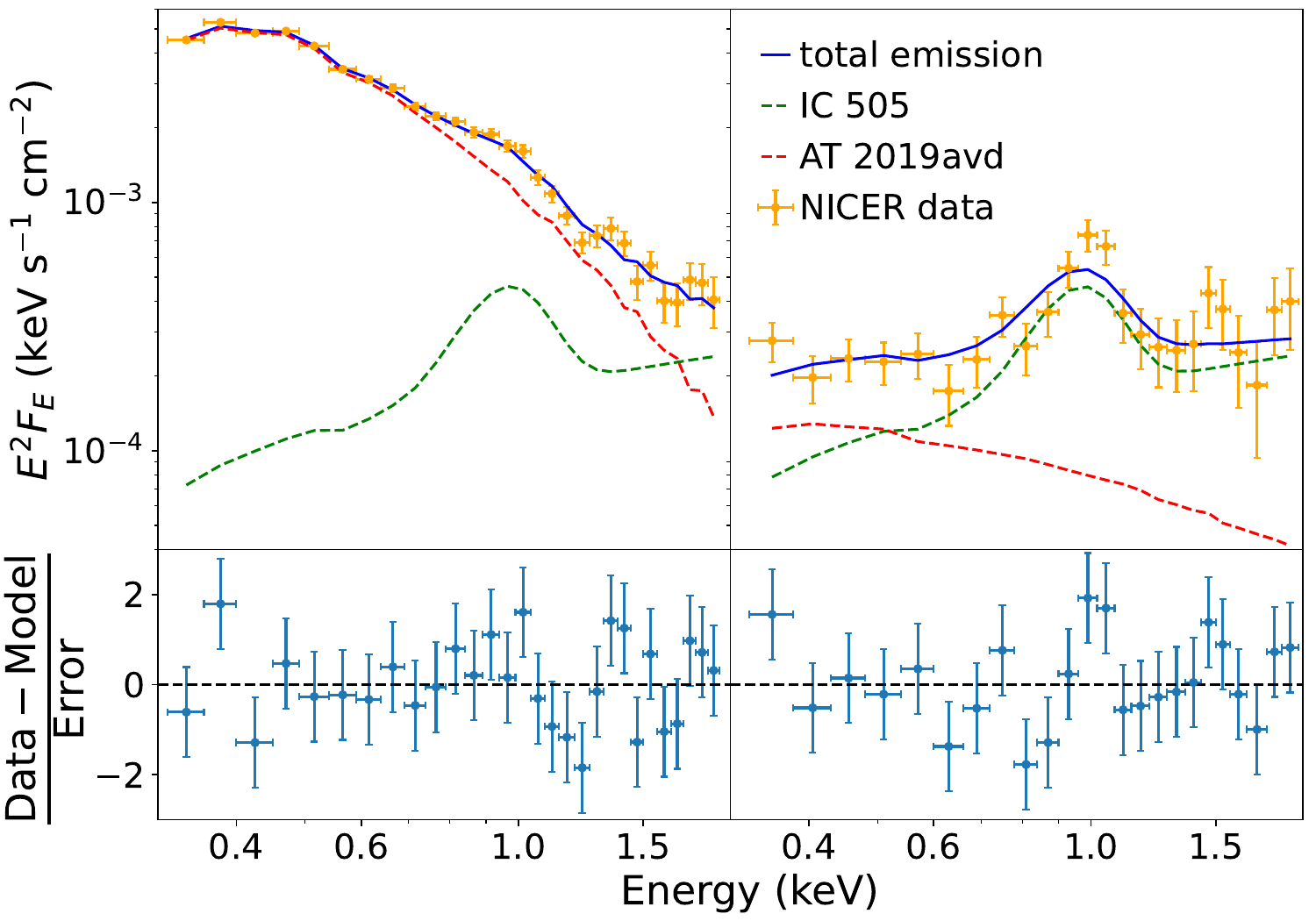}
    \caption{Representative unfolded X-ray spectra of AT 2019avd observed with \textit{NICER}, along with best-fitting model components and residuals. The left and right panels correspond to observations on Day 188 (Phase 3) and Day 244 (Phase 4), respectively. Residuals (bottom panels) are expressed in units of photon counts.} 
    \label{fig:spec_resid}  
\end{figure}

In our spectral analysis, we employed the \texttt{XSPEC} software to fit the \textit{NICER} spectrum of AT 2019avd using the model \texttt{TBabs*zashift*disk\_corona + IC 505}, where \texttt{disk\_corona} is a pre-computed table model generated using \texttt{Python}. The disc-corona system was assumed to be viewed face-on. Uncertainties are quoted
at 1$\sigma$ confidence level.
Figure~\ref{fig:spec_resid} shows two representative unfolded spectra and their corresponding residuals from Phases 3 and 4, respectively.

From Figure \ref{fig:spec_resid}, one can see that during Phase 3, the emission of AT 2019avd dominates across the full 0.3--2 keV band. However, at energies near and above 1 keV, the contribution from IC 505 becomes non-negligible and introduces substantial contamination. In Phase 4, IC 505 dominates almost the entire energy band.
Although the emission from IC 505 is relatively stable compared to AT~2019avd, which exhibits variations over two orders of magnitude over 1,000 days, IC 505 still shows modest variability of up to a factor of 3 (see the left panel of Figure A1 in \citealt{wangRapidDimmingFollowed2024}).
As a result, the use of a fixed IC 505 model across all spectra inevitably introduces systematic uncertainties, particularly near the overlapping regime around 1 keV and above. To mitigate this, we emphasize that only the residuals below 1 keV can reliably reflect the goodness of fit for the disc-corona model. For Phase 4, where the IC 505 component overwhelms the intrinsic emission of AT 2019avd, the derived disc-corona spectra must be treated with caution. In practice, we exclude a subset of spectra that show significant deviations from the general temporal trend of neighbouring observations, as these are likely influenced by discrepancies between the fixed IC 505 template and its actual emission during those epochs. For instance, the spectra obtained on Days 196.0 and 206.0 show significantly lower fluxes than those of adjacent epochs, while the spectrum on Day 211.2 appears unusually bright. In addition, the spectra on Days 235.2, 239.0, 243.1, and 248.5 are noticeably softer compared to their neighbouring observations, with the spectrum on Day 249.0 also showing a marginal softening. As reported by \citet{wangRapidDimmingFollowed2024}, AT2019avd exhibits significant variability on timescales of both hours and days. However, since our study focuses on the long-term evolution of AT2019avd on viscous timescales, such short-term fluctuations are not accounted for in our model. Overall, we include 8 \textit{NICER} observations in this work.

\begin{figure*}
    \centering
    \includegraphics[width=2.0\columnwidth]{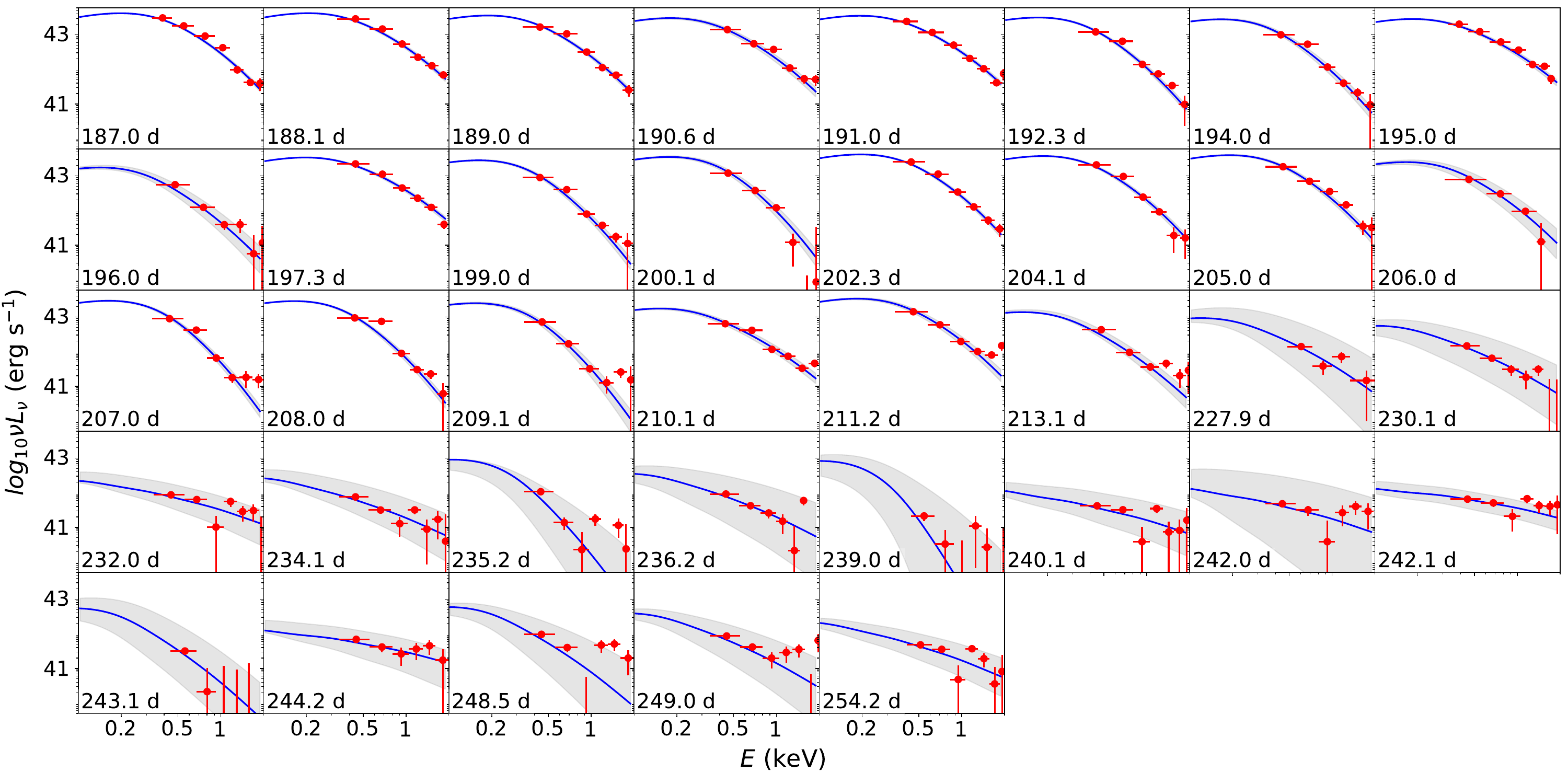}
    \caption{Time evolution of the intrinsic unabsorbed spectra of AT 2019avd in the 0.1--2~keV band, with emission from IC 505 subtracted. Each solid line represents the best-fitting spectrum at a given epoch and the shaded regions indicate the uncertainty range derived from the confidence intervals of the fitted model parameters.}
    \label{fig:spec_fit} 
\end{figure*}

In Figure~\ref{fig:spec_fit}, we present the unabsorbed X-ray spectra and fitting spectra of AT~2019avd in the 0.1--2~keV band, with the emission from IC~505 subtracted. Compared to Phase 3, the fitting uncertainties in Phase 4 (i.e., observations after Day 225.0) are substantially larger, primarily due to the increasing dominance of emission from IC~505.
Nevertheless, we can still identify a coherent evolutionary trend from the best-fitting model. In Phase 3, the peak of the accretion disc emission lies well within the 0.1--2~keV energy band and corresponds to a blackbody temperature of about 0.1~keV. The coronal emission is characterized by a very soft power-law. In Phase 4, the disc emission peak gradually shifts out of the observed energy band, and the X-ray emission becomes entirely dominated by the corona. In this phase, the luminosity below 1~keV decreases significantly relative to Phase 3, while at higher energy bands, there seems to be no noticeable decline in luminosity.

\begin{figure}
    \includegraphics[width=\columnwidth]{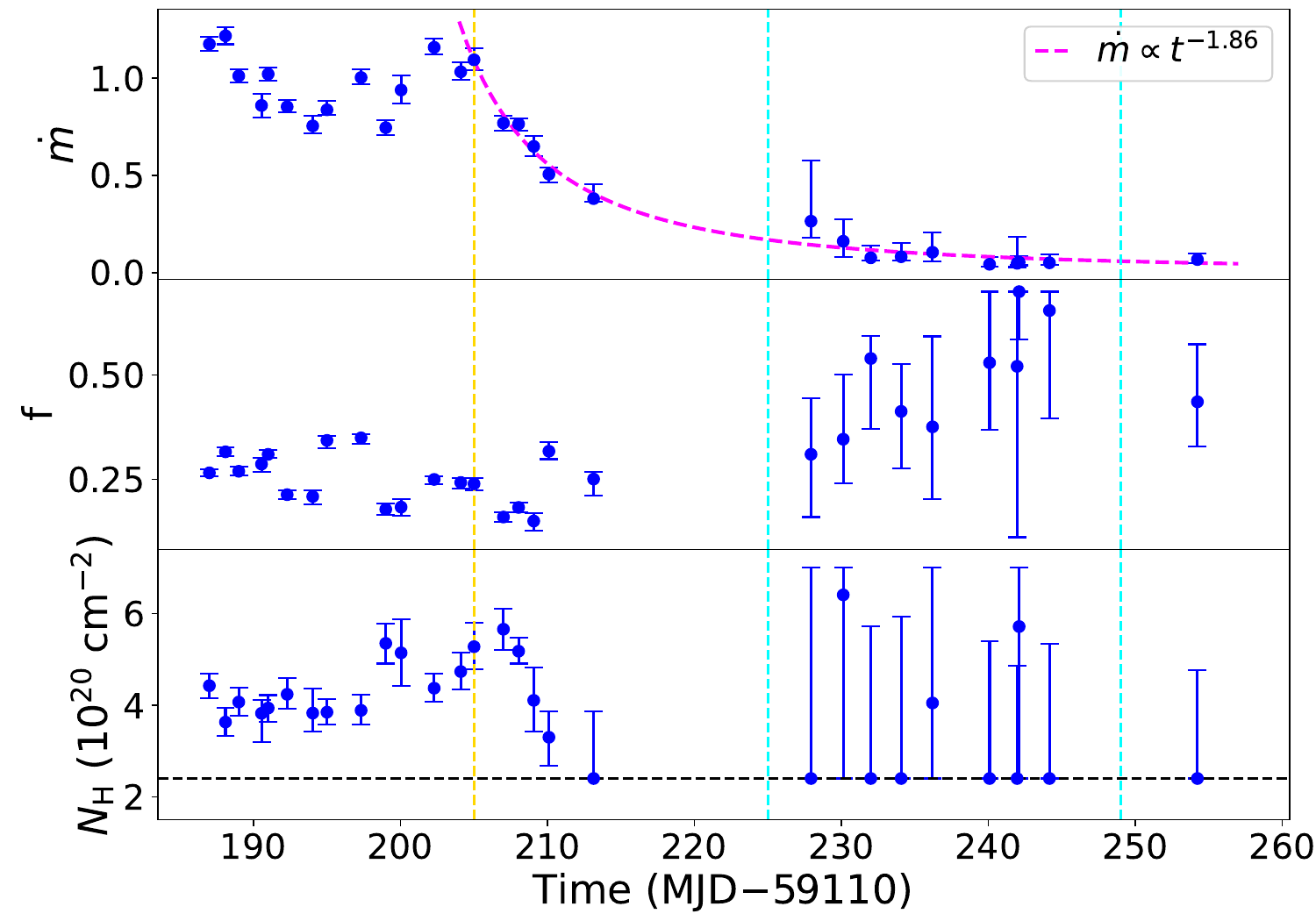}
    \caption{Temporal evolution of the best-fitting parameters for AT~2019avd. From top to bottom, the panels show: the accretion rate $\dot{m}$, coronal energy fraction $f$, and hydrogen column density $N_\text{H}$. The golden dashed line marks Day 205, corresponding to the onset of the rapid evolution phase. Cyan vertical dashed lines denote the start and end of Phase 4 (Days 224 and 249, respectively). In the top panel, purple and magenta dashed lines represent power-law fits to the declining segments of $\dot{m}$. In the bottom panel, the black dashed line indicates the lower bound of the fitted hydrogen column density.}
    \label{fig:fit} 
\end{figure}

In Figure~\ref{fig:fit}, we present the temporal evolution of the best-fitting parameters derived from our spectral modeling. The complete set of best-fitting parameters is tabulated in Appendix \ref{sec:appendix} (see Table \ref{tab:fit}). {The accretion rate $\dot{m}$ exhibits a sustained decay following approximately Day 205.}  
{To quantitatively constrain the decay rate, we adopt a generalized power-law form to fit the data from Day 205 onwards:}
\begin{equation}
    \dot{m} \propto {(t-t_0)}^{-\xi}, 
    \label{eq:mdot}
\end{equation}
{where $t_0 = 193.43 \pm 4.41~ {\rm day}$ and $\xi=1.86\pm0.47$ is the best-fitting index.}

Although subject to large uncertainties, the coronal fraction $f$ shows an overall increase from Phase 3 to Phase 4. During Phase 3, $f$ remains below 0.4, whereas in Phase 4, it predominantly exceeds this threshold, indicating a growing contribution from coronal emission as the system evolves. The best-fitting hydrogen column density, $N_{\rm H}$, remains relatively stable across the observations, with a mean value of approximately $5\times10^{20}{\rm cm}^{-2}$. However, a closer examination reveals a weak but systematic trend: $N_{\rm H}$ gradually increases prior to Day 205, followed by a mild decrease shortly thereafter. In some Phase 4 spectra, $N_{\rm H}$ reaches the lower bound of our fitting range, $2.4\times10^{20}{\rm cm}^{-2}$. Despite these modest variations, the overall stability of $N_{\rm H}$ lends confidence to the robustness of our derived accretion rate evolution.

We calculated the characteristic photon index $\Gamma$ in the high-energy band (0.6--2~keV), where the coronal emission dominates, using the best-fitting spectral model shown in Figure~\ref{fig:spec_fit}. This spectral slope reflects the power-law shape produced by inverse Compton scattering in the corona and should be free from contamination by disc emission. Using the best-fitting values of $\dot{m}$ and $f$, we reconstructed the corresponding coronal properties within the framework of our disc-corona model. Specifically, we extracted the electron temperature $T_{\rm e}$ and the optical depth $\tau$ at a representative radius of $13.5~R_{\rm g}$, where the radiative efficiency reaches its maximum.

\begin{figure} 
    \centering 
    \includegraphics[width=\columnwidth]{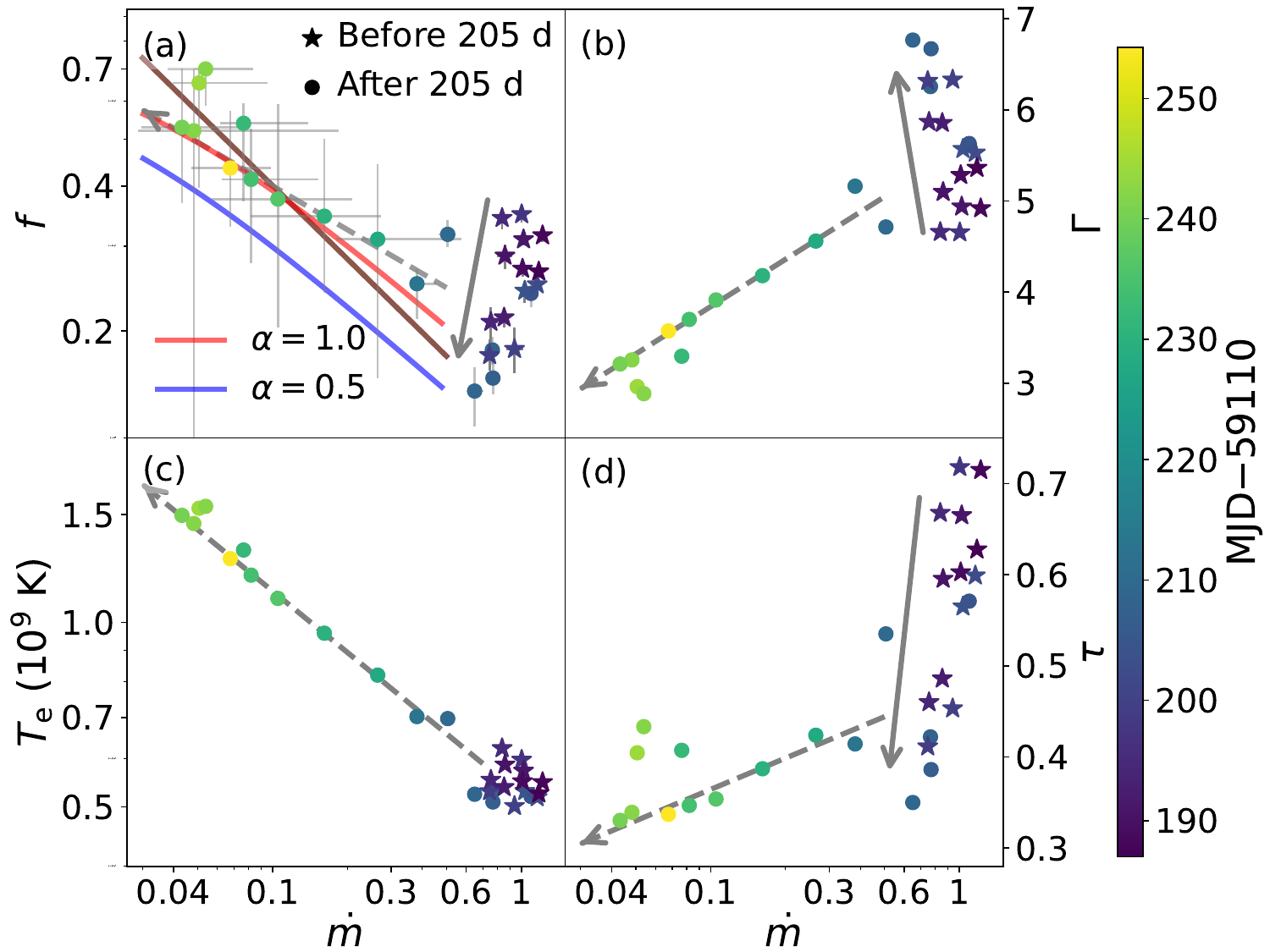}
    \caption{Evolution of coronal parameters for AT 2019avd as functions of the accretion rate $\dot{m}$. The four panels show: (a) the coronal energy fraction $f$, (b) the photon index $\Gamma$ in the 0.6--2~keV band, (c) the electron temperature $T_{\rm e}$ at $13.5~R_{\rm g}$, and (d) the optical depth $\tau$ at the same radius. Star-shaped markers represent results from before Day 205, while circular markers correspond to later observations. A colourbar encodes the observational time. Grey lines with arrows indicate the general evolutionary trends, and the dashed line indicates the best-fitting linear relationship for $\dot{m}\lesssim 0.3$. 
The blue and red solid lines in panel (a) represent the  model calculations of \citet{caoAccretionDisccoronaModel2009}, obtained with the viscosity parameter $\alpha=0.5$ and $\alpha=1$, respectively. {The brown solid line in panel (a) represents the best-fitting relation $L_{\rm X}/L_{\rm Bol}-L_{\rm Bol}/L_{\rm Edd}$, derived from our broadband extrapolation of the AGN data from \citet{wangHotDiskCorona2004} (see Section 5 for details).}}
    \label{fig:fit_paras}
\end{figure}

Figure \ref{fig:fit_paras} shows the evolution of the coronal energy fraction $f$, photon index $\Gamma$, electron temperature $T_{\rm e}$, and optical depth $\tau$ as functions of the dimensionless accretion rate $\dot{m}$. These parameters display distinct behaviours in different accretion regimes, suggesting a transition occurring in the corona as the source evolves. At higher accretion rates ($\dot{m}>0.6$), the parameter correlations are less clearly linear. A Spearman rank correlation analysis reveals statistically significant positive correlations for both $f$ and $\tau$ with $\dot{m}$, and a significant negative correlation for $\Gamma$ with $\dot{m}$. However, $T_{\rm e}$ shows no significant correlation with $\dot{m}$ in this regime.
Specifically, as $\dot{m}$ declines from its peak value towards about $0.6$, the coronal energy fraction $f$ decreases from approximately 0.3 to below 0.2, accompanied by an increase in the photon index $\Gamma$ from about 5 to nearly 7, indicating a softening of the X-ray spectrum. During the same period, the optical depth $\tau$ exhibits a substantial decrease from about 0.7 to about 0.4, suggesting a nearly 50\% drop in electron density within the corona, while the electron temperature $T_{\rm e}$ remains relatively constant. 
In Figure \ref{fig:fit_paras}(a), we also show the theoretical predictions for the $\dot{m}-f$ relation from the model of \citet{caoAccretionDisccoronaModel2009}, adopting different values of the viscosity parameter $\alpha$ for comparison.

At lower accretion rates ($\dot{m}<0.6$), the relationships between key coronal parameters and the accretion rate become approximately linear. Both $T_{\rm e}$ and $f$ increase significantly as $\dot{m}$ decreases, while the X-ray spectrum hardens ($\Gamma$ decreases). $\tau$ shows comparatively less variation but still exhibits a mild decreasing trend over time. When determining the best-fitting linear relations in this regime, we excluded the data points corresponding to Days 210.1, 232.0, 242.1, and 244.2, as these showed anomalously high values of f which could skew the trend analysis. 
The resulting best-fitting relations for $\dot{m}<0.3$ are  
\begin{align}
    &\log f = (-0.30\pm0.03) \log \dot{m} + (-0.70\pm0.03), \label{eq:fit_f} \\
    &\Gamma = (1.7\pm0.0) \log \dot{m} + (5.6\pm0.0), \label{eq:fit_Gamma} \\
    &\log T_{\rm e} = (-0.33\pm0.01) \log \dot{m} + (8.7\pm0.0), \label{eq:fit_Te} \\
    &\tau = (0.11\pm0.02)\log \dot{m} + (0.48\pm0.02). \label{eq:fit_tau}
\end{align}

When $\dot{m}$ falls to the lowest observed values around 0.04, $\Gamma$ approaches a value of about 3—still noticeably softer than the typical X-ray spectral slope of $\Gamma\approx2$ observed in most AGNs \citep[e.g.][]{2003AJ....125..433V}. Together, these results support a scenario in which the corona becomes hotter, relatively more dominant (higher $f$), and more optically thin as the accretion rate declines in AT 2019avd.

\section{Discussion}   
\label{sec:discussion}  

In this work, we utilized an accretion disc-corona model to investigate the X-ray spectral hardening observed in TDEs at later times. Our model successfully reproduces the X-ray spectral evolution of the TDE candidate AT 2019avd starting from 187 days after its peak emission.
Through our modelling, we derive the temporal evolution of key physical parameters, including the accretion rate $\dot{m}$ and the coronal fraction $f$. The results show that the accretion rate declines as $\dot{m} \propto (t - 193.43~{\rm day})^{-1.86}$ during the late stage. 
Moreover, we find a clear anti-correlation between $f$ and $\dot{m}$.
Our model also allows us to derive key coronal parameters, including the spectral photon index $\Gamma$, the coronal electron temperature $T_{\rm e}$, and the optical depth $\tau$. Our results show that when $\dot{m} \lesssim 0.6$, $\Gamma$, $T_{\rm e}$, and $\tau$ are strongly correlated with the accretion rate.
Specifically, as $\dot{m}$ decreases, $\Gamma$ increases, $T_{\rm e}$ rises, and $\tau$ drops slightly—indicating a progressively hotter and more tenuous corona that contributes more significantly to the X-ray emission. 
In the following, we discuss these results in detail, compare them with previous studies, and explore their broader implications for understanding accretion physics in BH systems.

Our model assumes that a fraction $f$ of the gravitational energy released in the accretion disc is transported into a hot corona, where it heats electrons that upscatter disc photons via inverse Compton scattering to produce the observed X-ray emission.  
For a given accretion rate $\dot{m}$, a lower value of $f$ implies less heating of the corona, resulting in a lower electron temperature $T_{\rm e}$ and a softer X-ray spectrum. When $\dot{m}$ decreases, the disc's radiative output decreases and cooling becomes less efficient, allowing the corona to remain hotter and produce harder spectra. This framework provides a natural physical explanation for the spectral hardening as the accretion rate decreases. 

{
In this model, the X-ray luminosity is approximately proportional to $\dot{m},f$. A decrease in $\dot{m}$ can be compensated by an increase in $f$, yielding a comparable X-ray flux. However, the X-ray spectral index is more sensitive to variations in $f$, so the spectral shape provides an additional constraint in the fitting process. A larger $f$ inevitably produces a harder spectrum, whereas a smaller $f$ results in a softer one. So, there is no parameter degeneracy in our model calculations. 
}

In AGNs, it has been well-established that key properties of the corona evolve with the accretion rate. For example, there exists a strong correlation between the Eddington ratio and the X-ray spectral index, $\Gamma$, where the spectrum softens as the accretion rate increases \citep[e.g.][]{shemmerHardXRaySpectral2006, shemmerHardXRaySpectrum2008}. Additionally, the coronal electron temperature, $T_{\rm e}$, inferred from the cut-off energies of AGN hard X-ray spectra, has been shown to decrease with an increasing Eddington ratio, reinforcing the picture that higher accretion rates lead to more efficient cooling in the corona \citep[e.g.][]{2018MNRAS.480.1819R}.

It is found that the anti-correlation between $f$ and $\dot{m}$ can be interpreted within the framework of magnetically heated corona models \citep[e.g.][]{caoAccretionDisccoronaModel2009}. \cite{caoAccretionDisccoronaModel2009} adopt the magnetic stress tensor form proposed by \cite{1984ApJ...287..761T},  $\tau_{r\varphi} = p_{\rm mag} = \alpha [p_{\rm gas}(p_{\rm gas}+p_{\rm rad})]^{1/2}$, to calculate the buoyant magnetic fields in the accretion disc and qualitatively explain both the $\dot{m} - f$ and $\dot{m}-\Gamma$ relations observed in AGNs. We compare the theoretical predictions for the $\dot{m}-f$ relation from the model of \cite{caoAccretionDisccoronaModel2009} with our results in Figure \ref{fig:fit_paras}(a) and find that the model can reproduce the observed $\dot{m}-f$ relation well if a viscosity parameter of $\alpha=1$ is adopted. Given that $\alpha<1$ is generally expected in accretion systems, this result implies a stronger magnetic field within the TDE accretion disc than that in typical AGNs. We speculate that this could be due to a stronger seed magnetic field from the disrupted star compared to that from the circum-nuclear gas, which is then amplified by the rotational instability (MRI) during the rapid formation of the transient disc.

The anti-correlation between the relative contribution of the corona and the accretion rate, analogous to our $\dot{m}-f$ relation, is also a well-established observational trend in AGNs. Studies consistently find a strong correlation between the ratio of X-ray to bolometric luminosity and the Eddington ratio, $L_{\rm X,2-10 keV}/L_{\rm bol}\propto (L_{\rm bol}/L_{\rm Edd})^{\xi}$, with the index $\xi$ typically close to $-0.6$ \citep[e.g.,][]{wangHotDiskCorona2004, 2012MNRAS.425..623L, 2020A&A...636A..73D}. It is found that both the photon index $\Gamma$ in the hard X-ray wave band and the black hole mass $m_{\rm BH}$ is available only in \cite{wangHotDiskCorona2004}'s sample. In order to compare their relation with the $\dot{m}-f$ relation derived in this work, we extract their data and extrapolate the 2-10 keV luminosity to a broader 0.1-100 keV band, assuming a single power-law, which serves as a better proxy for the total coronal power $Q_{\rm cor}$. Our re-analysis yields a relation $L_{\rm X} / L_{\rm Bol} \propto \left(L_{\rm Bol} / L_{\rm Edd}\right)^{-0.51}$ for their sample. Assuming $\dot{m}\simeq L_{\rm Bol} / L_{\rm Edd}$ and $f = L_{\rm X} / L_{\rm Bol}$, this empirical AGN trend is compared with the $\dot{m}-f$ relation derived for AT 2019avd in Figure \ref{fig:fit_paras}(a). 
The slope of the correlation between the coronal fraction and the accretion rate for AT 2019avd, $f\propto \dot{m}^{-0.30}$, during the later evolutionary phase when $\dot{m}\leq0.6$, is much shallower than that for the AGN sample. It is worth noting that previous studies of the $\dot{m}-f$ relation in AGNs have relied on large-sample analyses of different sources \citep[e.g.][]{wangHotDiskCorona2004, vasudevanPiecingTogetherXray2007}. Such approaches are subject to significant uncertainties arising from the determination of the BH mass in individual AGNs. In contrast, our study follows the evolution of a single system. The slope of the derived $\dot{m}-f$ relation is therefore independent of the BH mass estimate, and likely provides a more reliable probe of disc–corona coupling and evolution.

Our finding of late-stage spectral hardening with a declining accretion rate in AT 2019avd is consistent with a growing body of evidence from other TDEs, where similar rapid hardening has been observed on comparable timescales.
\citet{KOMOSSA2015148} pointed out that many TDEs exhibit spectral hardening over timescales of years, although earlier studies were often limited in characterizing this process due to the lack of continuous X-ray monitoring. More recently, similar rapid spectral hardening has been identified in TDEs with much more intensive observational coverage. For instance, \citet{weversLiveAnotherDay2023} and \citet{yaoMassiveBlackHole2025} reported rapid X-ray spectral hardening on timescales of roughly 30 days in TDEs {AT 2018fyk} and {AT 2024tvd}, which is remarkably consistent with our findings for AT~2019avd, where $\Gamma$ drops rapidly from $\sim5$ to $\sim3$ over a similar timescale. Such state transitions, moving towards harder spectra, are well documented in XRBs \citep[e.g.,][]{remillardXRayPropertiesBlackHole2006,wuXRaySpectralEvolution2008}, and have also been identified in AGNs, where the X-ray spectra harden as $\dot{m}$ declines \citep[e.g.,][]{wangHotDiskCorona2004, shemmerHardXRaySpectral2006, shemmerHardXRaySpectrum2008, risalitiSLOANDIGITALSKY2009}.

However, it seems that not all TDEs follow this evolutionary path \citep[e.g.,][]{auchettlNewPhysicalInsights2017, guoloSystematicAnalysisXRay2024}. 
The conditions for coronal formation and its subsequent strengthening appear complex. For instance, a systematic study by \cite{guoloSystematicAnalysisXRay2024} on optically selected TDEs found that the few cases in their sample showing spectral hardening tended to be associated with larger black hole masses. They observed that the corona became prominent primarily as accretion rates dropped to sub-Eddington levels. This finding is consistent with our results in the low accretion rate regime (i.e. $\dot{m}\lesssim 0.6$). However, \cite{guoloSystematicAnalysisXRay2024} also noted TDEs with similarly large black hole masses that did not show evidence of a significant corona, implying that other factors, potentially related to the magnetic field characteristics within the disc-corona system, play a crucial role in coronal development.

Intriguingly, it is also found that some TDEs with adequate X-ray monitoring but did not exhibit clear spectral hardening often showed evidence of powerful jets or outflows in the samples of \cite{auchettlNewPhysicalInsights2017} and \cite{guoloSystematicAnalysisXRay2024}. This leads to a preliminary hypothesis that such jet or outflow activity might suppress the formation of the hot corona, which is beyond the scope of this work. The underlying physical mechanism for this potential suppression is still unclear, and a larger sample with continuous multi-wavelength observations is needed to resolve this issue.

Additionally, while $\dot{m}>0.6$, the X-ray spectrum of AT~2019avd is dominated by the blackbody component, accompanied by an extremely {steep} power-law with photon indices $\Gamma>4$. 
Combined with the spectral hardening period, as the accretion rate declines from above 1 to around 0.04, the power-law component remains significantly steeper than those typically observed in AGNs or XRBs where the power-law indices are generally much lower (e.g., $\Gamma\sim1.5-2.5$). However, similar {steep} X-ray spectra have been reported in other TDEs \citep[e.g.,][]{badeDetectionExtremelySoft1996, komossaHugeDropXRay2004}. 
Given that the X-ray spectral hardness is highly sensitive to the values of $\dot{m}$ and $f$, it is plausible that the magnetic field processes are quite different between TDE and AGN discs.

As shown in Figure \ref{fig:fit}, the hydrogen column density, $N_{\rm H}$, exhibits a similar evolutionary trend to that of the accretion rate.
Before the rapid evolutionary phase around Day 205, $N_{\rm H}$ shows a mild but systematic increase. This could reflect the accumulation of absorbing material from various sources, including the inflowing streams of stellar debris themselves crossing our line of sight \citep[e.g.,][]{guillochonPS110jhDISRUPTIONMAINSEQUENCE2014, bonnerotDiscFormationTidal2016} or the development of a reprocessing layer or a puffed-up envelope characteristic of super-Eddington accretion phases \citep[e.g.,][]{strubbeOpticalFlaresTidal2009, metzgerBrightYearTidal2016}. 
The column density $N_{\rm H}$ drops sharply after Day 205, coinciding with the rapid decline of $\dot{m}$, consistent with a sudden dispersal or clearing of the surrounding or circumnuclear absorbing material.
Similar phenomena of rapid changes in $N_{\rm H}$ have also been found in other TDEs \citep[e.g.,][]{saxtonTidalDisruptionlikeXray2012, millerFlowsXrayGas2015, auchettlNewPhysicalInsights2017, weversBlackHoleMasses2019} and some changing look AGNs \citep[e.g.,][]{risalitiRapidComptonthickComptonthin2005, ricciChanginglookActiveGalactic2023}. 

Regarding the optical-to-UV emission, our disc–corona model predicts a luminosity of about $3\times 10^{40}~\rm erg~s^{-1}$, which is more than an order of magnitude lower than the value observed by \citet{wangRapidDimmingFollowed2024}. Moreover, \citet{wangRadioDetectionAccretion2023} estimated the optical-UV photospheric radius of AT2019avd to be $\sim 611 - 1256 R_{\rm g}$, significantly larger than the characteristic size of the accretion disc inferred in this work. These results suggest that the optical-to-UV emission in AT2019avd is unlikely to originate from the disc-corona system. This conclusion is consistent with the growing consensus that, in TDEs, emission from the accretion disc itself contributes significantly to the optical/UV band only at very late times, typically several years after the initial flare \citep[e.g.][]{vanvelzenLatetimeUVObservations2019}.

\section{Summary} 
In this work, we utilize a disc-corona model to explain the spectral hardening observed at later stages of TDEs, and we apply this model to the TDE candidate AT~2019avd. Our main results are summered as follows.
\begin{enumerate}
    \item An anti-correlation between $f$ and $\dot{m}$ was clearly observed when $\dot{m} \lesssim 0.6$. This trend is qualitatively consistent with that found in AGNs. The shallower slope ($f \propto \dot{m}^{-0.30}$) compared to that in AGNs ($f \propto \dot{m}^{-0.64}$) may suggest different magnetic field conditions between the two systems.
    \item The X-ray spectra harden as the accretion rate decreases, with $\Gamma$ dropping from 5 to 3 over approximately 30 days when $\dot{m} \lesssim 0.6$. Similar rapid spectral hardening has been observed in other TDEs, and the anti-correlation between $\Gamma$ and $\dot{m}$ is also observed in AGNs and BHXBRs.
    \item Derived from our model parameters, the corona is inferred to become progressively more dominant (higher $f$), hotter (increasing $T_{\rm e}$), and more tenuous (decreasing $\tau$) as the accretion rate declines in the low $\dot{m}$ regime. This provides a physical explanation for the observed spectral hardening.
    \item In the high accretion rate regime ($\dot{m} > 0.6$), the electron temperature of the corona ($T_{\rm e}$) remains nearly constant, while other parameters show different trends with the accretion rate. Whether this reflects a limitation of our model or a distinct physical behaviour (e.g., related to magnetic fields from the disrupted star) requires further investigation.
    \item Our fits reveal rapid changes in the neutral hydrogen column density ($N_{\rm H}$). This variability might be attributed to the accumulation of stellar debris streams or the evolution of a reprocessing layer/puffed-up envelope during super-Eddington accretion phases.
\end{enumerate}

\section*{Acknowledgements}

We are grateful to the referee for his/her insightful comments and suggestions. 
This work is supported by the NSFC (12533005, 12233007, 12347103, and 12361131579), the science research grants from the China Manned Space Project with No. CMS-CSST- 2021-A06, and the fundamental research fund for Chinese central universities (Zhejiang University). AAZ acknowledges support from the Polish National Science Center grants 2019/35/B/ST9/03944 and 2023/48/Q/ST9/00138.

\section*{Data Availability}

The data underlying this article will be shared on reasonable request to the corresponding author. 
The code used for the disc--corona model calculations, including the XSPEC table model and example implementations, is publicly available at
\url{https://github.com/XU-Haichao/at2019avd-disk-corona-model}. 



\bibliographystyle{mnras}
\bibliography{reference_mnras} 

\appendix
\section{Detailed Best-Fitting Parameters for AT 2019avd}
\label{sec:appendix}

\begin{table}
\centering
\renewcommand{\arraystretch}{1.3}
\begin{tabular}{ccccc}
    \toprule
    MJD-59110 &              $\dot{m}$ &                    $f$ & $N_{\mathrm{H}} / 10^{20}\,\mathrm{cm}^{-2}$ & $\chi^2 / \mathrm{dof}$ \\
    \midrule
        187.0 & $1.18_{-0.04}^{+0.04}$ & $0.27_{-0.01}^{+0.01}$ &                         $4.43_{-0.27}^{+0.27}$ &            $56.67 / 24$ \\
        188.1 & $1.22_{-0.04}^{+0.04}$ & $0.32_{-0.01}^{+0.01}$ &                         $3.63_{-0.31}^{+0.31}$ &            $25.44 / 23$ \\
        189.0 & $1.01_{-0.03}^{+0.04}$ & $0.27_{-0.01}^{+0.01}$ &                         $4.07_{-0.30}^{+0.31}$ &            $32.52 / 24$ \\
        190.6 & $0.86_{-0.06}^{+0.06}$ & $0.29_{-0.02}^{+0.02}$ &                         $3.82_{-0.63}^{+0.30}$ &            $43.38 / 23$ \\
        191.0 & $1.02_{-0.03}^{+0.03}$ & $0.31_{-0.01}^{+0.01}$ &                         $3.94_{-0.30}^{+0.28}$ &            $36.82 / 24$ \\
        192.3 & $0.85_{-0.03}^{+0.03}$ & $0.21_{-0.01}^{+0.01}$ &                         $4.24_{-0.31}^{+0.35}$ &            $39.31 / 24$ \\
        194.0 & $0.75_{-0.04}^{+0.05}$ & $0.21_{-0.02}^{+0.02}$ &                         $3.83_{-0.41}^{+0.53}$ &            $10.59 / 15$ \\
        195.0 & $0.84_{-0.03}^{+0.04}$ & $0.34_{-0.02}^{+0.01}$ &                         $3.85_{-0.27}^{+0.28}$ &            $15.17 / 16$ \\
        197.3 & $1.00_{-0.03}^{+0.04}$ & $0.35_{-0.01}^{+0.01}$ &                         $3.89_{-0.30}^{+0.34}$ &            $27.14 / 21$ \\
        199.0 & $0.75_{-0.04}^{+0.04}$ & $0.18_{-0.01}^{+0.01}$ &                         $5.35_{-0.44}^{+0.43}$ &            $29.77 / 25$ \\
        200.1 & $0.94_{-0.07}^{+0.08}$ & $0.18_{-0.02}^{+0.02}$ &                         $5.14_{-0.72}^{+0.74}$ &            $24.93 / 21$ \\
        202.3 & $1.16_{-0.04}^{+0.04}$ & $0.25_{-0.01}^{+0.01}$ &                         $4.37_{-0.29}^{+0.32}$ &            $22.91 / 23$ \\
        204.1 & $1.03_{-0.04}^{+0.05}$ & $0.24_{-0.01}^{+0.01}$ &                         $4.74_{-0.38}^{+0.41}$ &            $65.60 / 26$ \\
        205.0 & $1.09_{-0.05}^{+0.06}$ & $0.24_{-0.02}^{+0.01}$ &                         $5.28_{-0.50}^{+0.52}$ &            $16.80 / 24$ \\
        207.0 & $0.77_{-0.04}^{+0.04}$ & $0.16_{-0.01}^{+0.01}$ &                         $5.66_{-0.45}^{+0.45}$ &            $32.69 / 21$ \\
        208.0 & $0.76_{-0.04}^{+0.03}$ & $0.18_{-0.01}^{+0.01}$ &                         $5.18_{-0.27}^{+0.29}$ &            $30.09 / 21$ \\
        209.1 & $0.65_{-0.05}^{+0.05}$ & $0.15_{-0.02}^{+0.02}$ &                         $4.11_{-0.68}^{+0.72}$ &            $19.20 / 22$ \\
        210.1 & $0.51_{-0.04}^{+0.04}$ & $0.32_{-0.02}^{+0.02}$ &                         $3.30_{-0.62}^{+0.56}$ &            $50.09 / 27$ \\
        213.1 & $0.38_{-0.01}^{+0.07}$ & $0.25_{-0.04}^{+0.02}$ &                         ${2.40}_{{-0.00*}}^{+1.46}$ &            $23.08 / 24$ \\
        227.9 & $0.26_{-0.08}^{+0.31}$ & $0.31_{-0.15}^{+0.13}$ &                         ${2.40}_{{-0.00*}}^{{+4.60*}}$ &            $16.05 / 11$ \\
        230.1 & $0.16_{-0.08}^{+0.11}$ & $0.35_{-0.11}^{+0.16}$ &                         $6.41_{{-4.01*}}^{{+0.59*}}$ &            $22.97 / 23$ \\
        232.0 & $0.08_{-0.01}^{+0.06}$ & $0.54_{-0.17}^{+0.05}$ &                         ${2.40}_{{-0.00*}}^{+3.32}$ &            $30.91 / 23$ \\
        234.1 & $0.08_{-0.02}^{+0.07}$ & $0.41_{-0.14}^{+0.11}$ &                         ${2.40}_{{-0.00*}}^{+3.54}$ &            $26.70 / 20$ \\
        236.2 & $0.11_{-0.05}^{+0.10}$ & $0.38_{-0.17}^{+0.22}$ &                         $4.05_{{-1.65*}}^{{+2.95*}}$ &            $26.82 / 11$ \\
        240.1 & $0.04_{-0.01}^{+0.04}$ & $0.53_{-0.16}^{{+0.17*}}$ &                         ${2.40}_{ -0.00 *}^{+3.00}$ &            $20.74 / 23$ \\
        242.0 & $0.05_{-0.02}^{+0.14}$ & $0.52_{-0.41}^{{+0.18 *}}$ &                         ${2.40}_{-0.00*}^{+2.47}$ &            $33.22 / 18$ \\
        242.1 & $0.05_{-0.02}^{+0.03}$ & ${0.70}_{-0.11}^{{+0.00*}}$ &                         $5.72_{-3.32 *}^{+1.28 *}$ &            $23.22 / 23$ \\
        244.2 & $0.05_{-0.01}^{+0.04}$ & $0.65_{-0.26}^{{+0.05 *}}$ &                         ${2.40}_{-0.00 *}^{+2.93}$ &            $23.70 / 21$ \\
        254.2 & $0.07_{-0.02}^{+0.03}$ & $0.44_{-0.11}^{+0.14}$ &                         ${2.40}_{-0.00 *}^{+2.37}$ &            $42.06 / 26$ \\
    \bottomrule
    \end{tabular}
    \caption{Best-fitting parameters of the SED with the disc-corona model for AT 2019avd. Values marked with an asterisk (*) indicate parameters that reached a pre-defined boundary during the fit.} 
    \label{tab:fit}
\end{table}


\bsp	
\label{lastpage} 
\end{document}